# Personas as a Powerful Methodology to Design Targeted Professional Development Resources

Adrian Madsen, Sarah B. McKagan, American Association of Physics Teachers, College Park, MD 20740
Eleanor C. Sayre, Kansas State University, Manhattan, KS 66506
Mathew "Sandy" Martinuk, Alexander Bell, Cognition Technology, Vancouver, BC V5N3Z2
Email: {amadsen, smckagan}@aapt.org, esayre@ksu.edu, {sandy, alex}@cognitiontec.com

**Abstract:** Scaling and sustaining educational innovations is a common problem in the learning sciences. Professional development resources around educational innovations that are personalized to appeal to the needs and motivations of different types of faculty offer a possible solution. The method of developing "personas" to represent key types of users is commonly employed in user-interface design and can be used to produce personalized resources. Personas are fictional named archetypes of users encompassing generalizations of their key characteristics and goals that emerge from interviews. This method is especially powerful because personas succinctly package information into the form of a person, who is easily understood and reasoned about. Herein we describe the creation of a set of personas focusing on the needs and motivations of physics faculty around assessment and teaching. We present the personas, a discussion of how they inform our design and how the method can be used more broadly.

## Introduction

A common issue in the learning sciences is the scaling and sustaining of educational innovations. Educators and researchers continually focus on creating new curricula and teaching methods, but many promising innovations are not as effective when implemented by different teachers in new settings (Penuel, Fishman, Haugan Cheng, & Sabelli, 2011) so faculty stop using them (Henderson, Dancy, & Niewiadomska-Bugaj, 2012). The primary strategy used in science professional development at the university level is to disseminate educational innovations as finished products (Henderson, Finkelstein, & Beach, 2010). This strategy assumes that successful implementation does not depend on the details of the particular faculty members, students and institutions. It does not leave room for the faculty member to play a meaningful role in the improvements or take into account their unique environment (Henderson et al., 2010). This is problematic as faculty value the uniqueness of their personal styles, skills and preferences when choosing a teaching method (Henderson & Dancy, 2008). It may be that there is too much focus on getting faculty to use the educational innovation "as is" and not enough focus on the individuality of the faculty and their environments.

One way to address scale and sustainability is through personalized professional development designed to take into account the diverse needs, motivations and environments of teachers and faculty who are potential or current users of educational innovations. We believe that personalized professional development resources will increase implementation and continuing use of these innovations because they target the things that faculty members actually care about. We are specifically designing a professional development website, the Physics Education Researcher (PER) User's Guide (www.perusersguide.org), to support faculty use of curricular and assessment innovations in physics at the university level. Our aim is to understand the needs, motivations and environments of different types of faculty, prioritize which types of faculty our site will be designed for and design our website in a way that speaks to each and meets real needs.

We propose that the method of creating "personas" is a powerful way to understand different types of faculty members who are potential or continuing implementers of educational innovations. This method is primarily used in the user-interface design process where designers create new technology products to match the needs of potential users (Cooper, 2004; Jacobs, Dreessen, & Pierson, 2008). The creation and use of personas is a well-established research technique where rich sets of qualitative data about users' goals and experiences are synthesized into user archetypes called "personas." These personas are generalizations of key characteristics and goals of potential users. Personas are described as "fictional, detailed archetypical characters that represent grouping of behaviors, goals and motivations observed and identified during the research phase" (Calde, Goodwin, & Reimann, 2002). They are usually named and assigned a picture to help them feel like real people. Personas as a method has roots in ethnography (Blomberg & Burrell, 2009) as well as human computer interaction (Sears & Jacko, 2007).

A common problem in the design process (we use the term "design" in the engineering sense) is having too vague an idea of who the users are. Without clarity, it is impossible to communicate about the specific needs and goals of the users and design a resource or experience to meet these. This is referred to as the problem of the "elastic user" who stretches and changes as the needs and constraints of the project change (Cooper, 2004). To address this problem, one could design for a specific user, but this produces a product that is too narrowly focused (Jacobs et al., 2008). Instead one could compile information on many users in the form of a report on key demographics and characteristics of users. The detailed nature of these reports makes them cumbersome to use (Pruitt & Adlin, 2010) and they are not naturally reasoned about by the human mind (Grudin, 2006).

In order to help designers understand, communicate about and effectively design for a variety of users, personas are used. These clearly defined representations of people package a large amount of information into a succinct format that is easy to understand and gives the design team something concrete to refer to and discuss (Pruitt & Adlin, 2010; Yström, Peterson, Von Sydow, & Malmqvist, 2010). Personas help the designers become user-centered and avoid self-centered designs (Pruitt & Adlin, 2010). Personas are an especially powerful communication tool because we as humans are innately equipped to generate and engage with representations of people, as opposed to statistical summaries, which are more difficult for us to engage with (Grudin, 2006). The personas do not represent individual users but contain key differentiating characteristics from an amalgamation of users. This allows designers to focus on important characteristics and not get sidetracked by superfluous details. The abstraction of personas away from the interviewees upon which they are based also ensures the anonymity of participants, which is especially important when focusing on sensitive issues that participants would not discuss unless anonymity were guaranteed. Further, there are cases where even though participants are not identified outright, their uniqueness creates an anonymity set too small to protect their identity. Creating personas for these kinds of participants ensures their protection. Finally, personas are referred back to throughout the entire project to ensure that the product does not depart from the needs of the actual users.

There are several limitations of the method of personas. Because personas blend together characteristics of many users, it is hard to tell which and how many users each persona actually represents (Chapman & Milham, 2006). There is a propensity for designers to base their understanding of users on stereotypes and not the actual users (Turner & Turner, 2011). This can be mitigated by closely coupling data to the persona creation. Personas are difficult to validate, as they emerge from an interaction of the specific set of users interviewed, the questions discussed and decisions by the design team (Chapman & Milham, 2006). The designers make conscious decisions about what aspects of the users to include in the personas based on which aspects they understand are important to the final product. It is likely that given a similar topic but different set of users and designers, different personas would emerge. That being said, validity can be established with expert review (Yström, Peterson, Sydow, & Malmqvist, 2009). The effectiveness of the final product for the intended user base is an additional measure of validity.

In this paper we detail our methodology, describe the personas that emerged from our interviews and discuss their usefulness in designing professional development resources around educational innovations.

## Methodology: Creation of Personas

We used semi-structured interviews to learn about faculty members' and department heads' current practices and needs around assessment in physics. We interviewed 13 physics faculty and 11 physics department heads in online interviews. Video was used in most. We recruited faculty members from a list of participants from a professional development workshop for new faculty held several years earlier. We sent faculty an email invitation to participate. We subsequently received recommendations from current participants and the project staff for other faculty who might be willing to participate and invited them. We recruited department heads from a list of participants from a different professional development workshop held the previous summer. A sufficient number of these department heads agreed to participate, so no additional invitations were needed. Our overall sample included eight participants from undergraduate serving public institutions, six from undergraduate serving private institutions and 10 from research institutions.

As part of our interview protocol, we asked participants about their background, school and department, current teaching practices and use of assessment, needs around assessment and how our online resources might meet their needs. Our interview team included two user interface designers (one is a former physics education researcher) and a physics education researcher. During most of the interviews, one team member engaged with the interviewee while the others listened and took notes. The interviews were recorded and after each interview, the team members individually wrote down the key points they noticed, primarily attending to the user's motivations and goals, tasks that they commonly completed around assessment, attitudes and beliefs, needs, pain points and constraints. These items were often discussed amongst the team members before the next interview. After several interviews, the team identified a few characteristics that varied among the interviewees and strongly influenced their teaching and assessment practices. For example, the interviewees' level of "buy-in" to educational innovations and knowledge of these resources varied significantly and had a strong impact on their use of research-based materials. We plotted the participants on these two axes and found groupings of participants, which helped in the initial development of the personas. As the interviews progressed, we continued to identify key differentiators and used a constant comparison protocol to determine if they were meaningful. As we reached saturation, we developed an initial profile for each persona and tested these to see if they fit the actual users we had interviewed.

After we had completed all the interviews, one team member re-listened to most of the interviews and took additional notes on goals, motivations and needs, writing down key illustrative quotes. These quotes were used to create a document synthesizing the tasks, behaviors and attitudes of each persona. These quotes and synthesis were used to once again modify the personas. Finally, we discussed the current version of the personas as a team and made additional changes based on evidence presented in the discussion.

Using the final versions of the personas, we ranked each on the variety of differentiating characteristics identified during the persona construction process, e.g., the degree to which faculty valued evidence or intuition to guide their teaching and assessment practices. These rankings helped fill out the important characteristics of the personas. With a rich understanding of the personas, we created task flows and outlined associated pain points. Task flows are detailed lists of steps describing common tasks users complete around the aim of the project, in our case, common tasks around teaching and assessment. Examples include finding a new research-based assessment or teaching method or comparing the results of an assessment. Within each task flow, we identified "pain points" discussed by users. Pain points are any inconvenience, confusion or frustration that users encounter when performing a specific task, for example, uncertainty on how to interpret the results of an assessment or difficulty finding appropriate data to compare one's results to.

Finally, the design team met with the larger group of project stakeholders to prioritize the personas with respect to the design of the website. They discussed their own goals for the website and the likelihood that a particular persona would benefit from the site's offerings in order come to a consensus decision on which personas to design the site for.

## Results: The Personas

Five personas emerged from our interviews (Table 1). In order to discuss these personas most naturally, they were given names and assigned pictures. We also chose key quotes from the interview transcripts to help us grasp who each persona represented and outlined their goals and motivations as related to assessment and teaching. Table 2 contains rankings of the personas on key characteristics that differentiate them from one another. Tables 1 and 2 are used together to get a full picture of what constitutes each persona. The demographics of the personas are not meant to be a meaningful characteristic. We recognize that these do not reflect the demographic distribution of US physics faculty.

After creating the personas, we outlined the key task flows and pain points that each would have when interacting with our online resources. For example, a common task flow for Diane, the pragmatic satisficer, is to search for a research-based assessment, run the assessment, analyze and interpret the results and compare the results with others. Her pain points include a concern that students are not taking it seriously, too much time is required to analyze the results and it is difficult to find data from other institutions for comparison. Several other task flows and associated pain points emerged for Diane during the interviews. This information gives the team a rich sense of who Diane is and her wants and difficulties, which allows us to design our online resource to meet her stated needs as well as infer solutions to other needs that users like her may not have voiced.

After review and discussion of the personas and our own goals for the site, the stakeholder team prioritized who the site will be designed for. We believe that we can have the most impact on scaling and sustaining educational innovations if we provide resources to faculty who are already motivated to improve student learning. We feel our impact would be lessened if we focused on convincing skeptical faculty to use these innovations. Further, faculty who are already successful at using innovations don't need our help as much as those with little experience. So we prioritized in order from most to least important, Raphael the motivated novice, Tim the seeker, Diane the pragmatic satisficer, Marge the proto-researcher and Paula the skeptic.

## Analysis and Discussion

Understanding who we should design for, their task flows and pain points, and prioritizing these personas tells us a lot about how to design our online resources so they are targeted and personal. For example, different personas thrive with different levels and kinds of information about educational innovations. Raphael is brand new to using educational innovations and doesn't have a lot of time, so he needs simple practical information and ready-to-go teaching and assessment materials. He also needs resources on how to face obstacles he will encounter when using these new methods which will support him in persisting in his use of the innovations. On the other hand, Tim has experience using educational innovations and wants lots of details about the research-validation and theoretical underpinnings of the innovation. He would also like information on how this innovation compares to other innovations he is familiar with. This level of detail will help him intelligently choose the most appropriate innovation for his own environment. Paula has a low buy-in to the value of educational innovations but cares about student learning, so she needs resources to help her understand how educational innovations can improve learning before being interested in the details of a new innovation.

There is rarely a professional development situation where only one persona is represented. Further, an individual faculty member may have characteristics from several personas. To address this, different aspects of the professional development resources should be designed to appeal to the personas based on the prioritization decided on by the stakeholder team. This is in contrast to the common practice of providing one type of professional development resource that might appeal to only one persona or contain so little or so much information that it benefits no one at all. We briefly discuss how we personalized our website based on the personas. The needs of Raphael were our top priority, so we designed a simple home page with very basic but enticing information and a jump bar with links to articles answering his commonly asked questions. From the home page, we provided links to more detailed pages so that Diane can, for example, compare her students to others using a database of assessment results to get the number she cares about. We also provided links from the home page for Tim leading him to detailed information about the research validation for each innovation and recommendations for new innovations to try. We believe design choices like these will help faculty who identify with a given persona easily find what they need to implement or continue using educational innovations.

We believe the method of personas is useful whenever researchers seek to design a professional development resource or experience aimed at scaling or sustaining educational innovations. Personas encourage the design team to focus on the users and transform a large amount of detailed information about individuals into digestible outlines of fictional people who can be easily be understood and discussed. This method also provides a framework for how to design professional development targeted at different types of users and fills in the details of who the users are and what needs or motivations they have that the innovation can satisfy. With an increased focus on the users and not just the innovations themselves, we believe more educators will take up new innovations and continue to use them.

Table 1: Descriptions of five personas emerged from interviews with physics department heads and faculty. These personas represent user archetypes, but no persona exactly represents any one user. (names and pictures are included to help communicate about the personas as if they were real people)

| | Role | Key Quote | Motivations and Goals |
|---|---|---|---|
| Paula the Skeptic 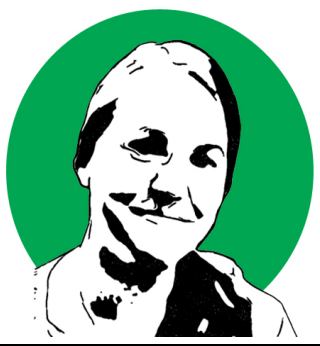 | 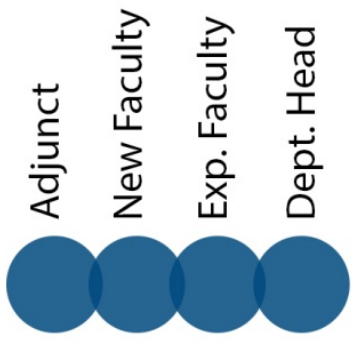  | "The entire assessment movement seems to ignore that we are already assessing our students." | • Unconvinced that research-based techniques offer anything more than traditional/current methods<br>• Assessment to reaffirm validity of current teaching methods<br>• External pressures for assessment |
| Raphael the Motivated Novice 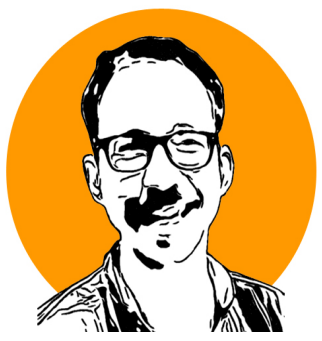 | 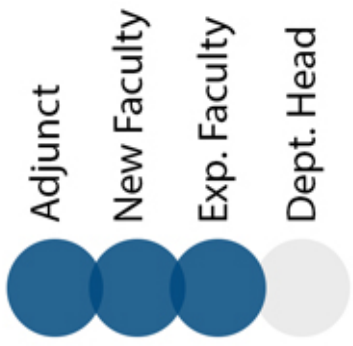  | "... make it useful for the students. I mean at the end of the day, that's all I really care about." | • Thoughtful teacher who cares about student learning<br>• Traditional background or new to teaching<br>• Acknowledges challenges with traditional methods<br>• Time-pressured |
| Diane the Pragmatic Satisficer 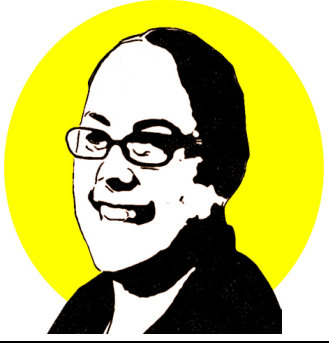 | 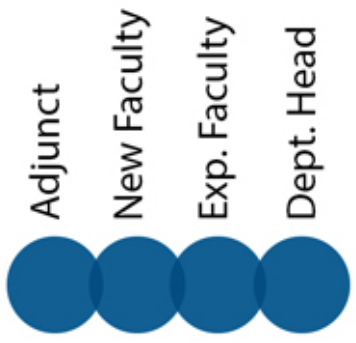  | "The ability for faculty to compare what they are doing with other faculty would be really useful and spur them on to improve their teaching." | • Interested in the result of assessment: the number<br>• Gives assessment to compare results<br>• Wants to know what teaching methods "work"<br>• Prefers summary information |
| Tim the Seeker 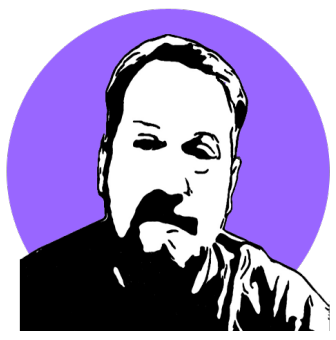 | 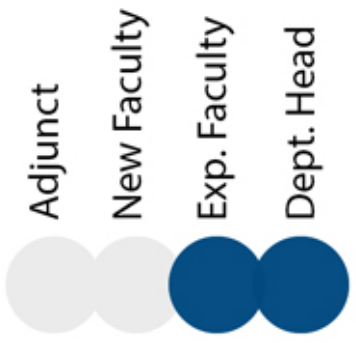  | "We're driven by data & cause & effect. If the data shows that the students are getting more out of the course, then that gives you reason to change." | • Generally curious and interested in student learning improvement<br>• Asks questions which lead to change/ improvement<br>• Wants to find suitable existing tools<br>• Internally motivated |
| Marge the Proto-researcher | 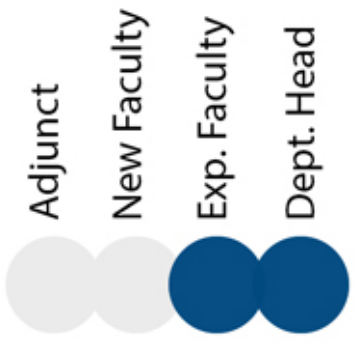  | "It's hard to argue that you don't want data about what your students are learning in your class." | • Has questions or needs beyond those already answered by educational research<br>• Wants to assess new aspects of students understanding / development<br>• Needs to do more analysis of results |
| 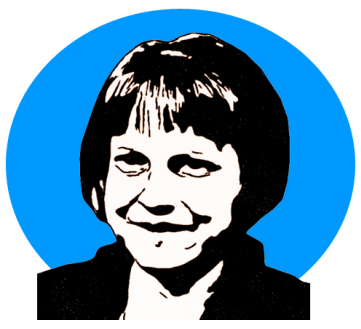 | | | |

Table 2: Ranking of personas on characteristics that emerged in user interviews. P = Paula the Skeptic, R = Raphael the Motivated Novice, D = Diane the Pragmatic Satisficer, T = Tim the Seeker, M = Marge the Proto-researcher, PER = physics education research.

| | | | | | | |
|---|---|---|---|---|---|---|
| High PER knowledge | M | T | D | R | P | Low PER knowledge |
| High PER buy-in | T M | D | R | | P | Low PER buy-in |
| Reformed teacher | M | T | D | R | P | Traditional teacher |
| Evidence-based | M | T | D | R | P | Intuition-based |
| Teacher responsibility for learning | | T M | D | R | P | Student responsibility for learning |
| Cares about meaning of results | M | T | R P | | D | Cares about numerical result |
| Seeks expert opinion | D M | T | | | P | Confident in own opinion/skills |
| High influence over teachers | M | T | | D P | R | Low influence over teachers |